\documentclass[10pt]{iopart}
\newcommand{\csch}{\mathrm{csch}}
\usepackage{cite}
\usepackage{bm}
\usepackage{graphicx}
\usepackage{cuted}
\newcommand{\TE}{\textsf{TE}}

\newcommand{\TM}{\textsf{TM}}
\newcommand{\CP}{\textsf{\scriptsize CP}}
\newcommand{\Ca}{\textsf{\scriptsize C}}
\newcommand{\te}{\textsf{\scriptsize te}}
\newcommand{\tm}{\textsf{\scriptsize tm}}

\newcommand{\adb}{} 
\newcommand{\ann}{\nonumber \\}
\newcommand{\cF}{\mathcal{F}}

\usepackage{color} \definecolor{darkgreen}{rgb}{0,.5,0}
\usepackage[colorlinks,filecolor=blue,citecolor=darkgreen,unicode]{hyperref}
\begin{document}
\title[Thermal Casimir and Casimir-Polder interactions]{Thermal Casimir and Casimir-Polder interactions in $N$ parallel 2D Dirac materials}
\author{Nail Khusnutdinov$^{1,2}$, Rashid Kashapov$^2$ and Lilia M. Woods$^3$}
\address{$^1$ Centro de Matem\'atica, Computa\c{c}\~ao e Cogni\c{c}\~ao, Universidade Federal do ABC, 09210-170 Santo Andr\'e, SP, Brazil}
\address{$^2$ Institute of Physics, Kazan Federal University, Kremlevskaya 18, Kazan, 420008, Russia}
\address{$^3$ Department of Physics, University of South Florida, Tampa, Florida 33620, USA}
\ead{nail.khusnutdinov@gmail.com}
\begin{abstract}
The Casimir and Casimir-Polder interactions are investigated in a stack of equally spaced graphene layers. The optical response of the individual graphene is taken into account using gauge invariant components of the polarization tensor extended to the whole complex frequency plane. The planar symmetry for the electromagnetic boundary conditions is further used to obtain explicit forms for the Casimir energy stored in the stack and the Casimir-Polder energy between an atom above the stack. Our calculations show that these fluctuation induced interactions experience strong thermal effects due to the graphene Dirac-like energy spectrum. The spatial dispersion and temperature dependence in the optical response are also found to be important for enhancing the interactions especially at smaller separations. Analytical expressions for low and high temperature limits and their comparison with corresponding expressions for an infinitely conducting planar stack are further used to expand our understanding of Casimir and Casimir-Polder energies in Dirac materials. Our results may be useful to experimentalists as new ways to probe thermal effects at the nanoscale in such universal interactions.
\end{abstract}  
\submitto{TDM}
\ioptwocol
\maketitle

\section{Introduction}

Long-ranged dispersive forces originating from zero-point vacuum fluctuations exist between any types of objects. Specifically, the presence of boundaries and/or molecular structures modifies the electromagnetic boundary conditions, which in turn gives rise to such fluctuation induced forces including Casimir and Casimir-Polder interactions \cite{Casimir:1948:otabtpcp,Casimir:1948:tiorotlvdwf,Lifshitz:1956:ttomafbs}. Dispersive forces dominate in inert materials and they are related to important phenomena, such as stability of composites, adhesion, and sticktion in tiny instruments and biological matter among others \cite{Woods:2016:mpocavdwi,Klimchitskaya:2009:tcfbrmeat}. Understanding the underlying mechanisms is important not only for the fundamental science of these ubiquitous forces, but also for control in the laboratory. The Casimir-Polder interaction, in particular, is especially strong in near-field interference setups, where distortion of the wave fronts due to different atoms or surfaces can lead to significant changes \cite{Nimmrichter:2008:tonfmwibtea,Gerlich:2007:akifhpm}. Trapping atoms near surfaces of Bose-Einstein condensates are also sensitive to Casimir-Polder forces  \cite{Leanhardt:2003:becnams,Druzhinina:2002:eooqrithel}. The Casimir force, on the other hand, can limit the operation of nano/micro electro-mechanical systems or it can be used as a driving force for actuators \cite{Chan:2001:qmaomsbtcf,Buks:2001:saeatceims,Esquivel-Sirvent:2010:vdwtibemf}.

Recently, special attention has been devoted to long-ranged dispersive forces involving carbon materials, such as graphene, carbon nanotubes and nanoribbons \cite{Drosdoff:2014:qatdfatgn,Popescu:2011:cdcni,Bordag:2006:ltffgaswcnvdwaci,Sarabadani:2011:mbeitvdwibgl,%
Gomez-Santos:2009:tvdwibgl}. The Dirac-like electronic structure and the unique optical properties of these materials have resulted in new functionalities in terms of magnitude, sign, and characteristic distance dependences, which are not found in Casimir interactions in standard materials \cite{Woods:2016:mpocavdwi,Klimchitskaya:2009:tcfbrmeat}. Moreover, it has been shown that the Hall phase diagram in the graphene family, composed of graphene, silicene, germanene, and stanene, accessible via external fields results in Casimir force phase transitions with even greater range of tunability \cite{Rodriguez-Lopez:2017:cfptitgf}. With experiments being performed at room temperature, the impact of thermal fluctuations on Casimir-Polder/Casimir phenomena is of special interest. It appears that the massless charge excitations in 2D graphene materials have strong effects on dispersive interactions when temperature is taken into account. Specifically, due to the much smaller Fermi velocity compared to the speed of light, the onset of thermal fluctuations involving 2D Dirac-like materials begins at much smaller separations as compared to typical metals and dielectrics \cite{Sushkov:2011:oottcf,Bimonte:2017:htotgteitcffgs}.   

The remarkable manifestation of thermal fluctuations in graphene materials is intertwined with their electronic structure and optical response properties. A successful model for the graphene response properties is the polarization tensor \cite{Bordag:2015:qftdftrog,Klimchitskaya:2016:copgtautpt}. This approach is based on the graphene Dirac Hamiltonian and it takes into account temperature and spatial dispersion through the wave vector. This polarization tensor description is complementary to the recently presented optical conductivity tensor for the entire graphene family based on the linear response Kubo formalism \cite{Rodriguez-Lopez:2018:noritptitgf}. The reflection coefficients of graphene have been expressed in terms of the polarization tensor and have been utilized in the Lifshitz approach to calculate Casimir interactions involving systems with a single graphene layer \cite{Bordag:2015:qftdftrog}. It has been demonstrated that the strong thermal effects at small separations originate not only from the zero Matsubara dominance, but also from the polarization tensor dependence on temperature, frequency, and wave vector \cite{Klimchitskaya:2015:oolteitcibtgs}. These strong thermal effects cannot be captured by other, simpler models for the graphene response properties, such as a constant conductivity model \cite{Kuzmenko:2008:uocog}, which reflects the small frequency range (less that $3$ eV) and is thus suitable for Casimir interactions at large distances \cite{Drosdoff:2010:cfags,Fialkovsky:2012:frig}. 

In this work, we consider the Casimir interaction energy stored in a stack of $N$ graphene planes and the Casimir-Polder interaction between an atom and the layered graphene stack. Due to the planar symmetry of the system, the response of each graphene can be represented as two decoupled conductivities, which are expressed in terms of the gauge invariant components of the polarization tensor extended to the whole complex frequency plane. This enables the numerical evaluation of the different interactions as a function of separation, temperature, mass gaps, and chemical potential. Results obtained via the constant conductivity show how the two models for the response compare in different regions. Numerical and analytical results for a stack of infinitely conducting planes are also shown as a reference and further comparisons. This work is especially beneficial as it provides a unified approach for Casimir and Casimir-Polder interactions in configurations with finite number of infinitely thin layers by taking into account the most sophisticated and complete analytical representation of the graphene response. Studying more than one graphene layers and tuning the chemical potential are especially interesting from experimental point of view, since this can be done by changing $N$ and varying the external electrostatic potential. Our results offer new opportunities for probing Casimir interactions and micro- and nanomechanical device applications.

\section{Optical response} 
The Casimir and Casimir-Polder interactions are determined by the electromagnetic modes supported by the system. For planar systems, which are of interest here, these can be separated into transverse magnetic (\TM) and transverse electric (\TE) contributions. The optical response of the individual graphenes in the multilayered system is also a key ingredient for the interactions. A reliable approach here is to utilize the polarization tensor approach, which enables taking into account the electronic structure of graphene as well as frequency, temperature, and chemical potential \cite{Bordag:2009:cibapcagdbtdm,Fialkovsky:2011:ftcefg,Bordag:2016:ecefdg,Bordag:2017:eecefdg}. The components of the polarization tensor, $\Pi_{lj}$, can be related to the components of the optical conductivity tensor $\sigma_{lj}$ as
\begin{equation}
\sigma_{lj} = \frac{\Pi_{lj}}{i\omega}.
\end{equation}
Due to gauge invariance, the polarization tensor has only two independent components, $\Pi_{00}$ and $\Pi_{tr} = \Pi^{\mu\nu} g_{\mu\nu} = \Pi_{00} - \Pi_{11} - \Pi_{22}$. The explicit expressions for $\Pi_{00}$ and $\Pi_{tr}$ were previously obtained in Ref. \cite{Bordag:2009:cibapcagdbtdm,Fialkovsky:2011:ftcefg,Bordag:2016:ecefdg,Bordag:2017:eecefdg}. It turns out that the graphene optical conductivity tensor entering into the reflection coefficients for the interactions calculations can be decoupled into components related to the gauge invariant quantities  $\Pi_{00}$ and $\Pi_{tr}$. Specifically, using imaginary frequencies $\lambda = i \omega$ the normalized to the graphene universal conductivity $\sigma_{gr} = e^2/4$, the $\overline{\sigma}_\tm$ and $\overline{\sigma}_\te$ components, corresponding to the TM and TE modes, are expressed as 
\begin{equation}
\overline{\sigma}_\tm = \frac{4\lambda}{e^2k^2} \Pi_{00}, \overline{\sigma}_\te = \frac{4}{e^2\lambda}\left(\Pi_{tr} -\frac{\lambda^2+k^{2}}{k^2} \Pi_{00}\right),
\end{equation}
where the trace of the $\overline{\bm{\sigma}}$ tensor is $\Tr \overline{\bm{\sigma}} = \frac{4}{e^{2} \lambda} \left(\Pi_{tr} - \Pi_{00}\right) =  \frac{4}{e^{2} \lambda} \Pi^k_k$. Here and throughout the paper all relations are given in $\hbar = c = k_B =1$ units. 

The expressions for the two types of conductivities can further be rewritten as 

\begin{strip}
\begin{eqnarray}
\overline{\sigma}_\tm &=& \frac{4m\lambda }{\pi(\lambda^2 + v_F^2 k^2)}\left(1 + \frac{\lambda^2 + v_F^2 k^2 - 4m^{2}}{2m\sqrt{\lambda^2 + v_F^2 k^2}} \arctan \frac{\sqrt{\lambda^2 + v_F^2 k^2}}{2m}\right)  + \Delta\overline{\sigma}_\tm,\ann
 \overline{\sigma}_\te &=& \frac{4m}{\pi \lambda} \left(1 + \frac{\lambda^2 + v_F^2 k^2 - 4m^{2}}{2m\sqrt{\lambda^2 + v_F^2 k^2}} \arctan \frac{\sqrt{\lambda^2 + v_F^2 k^2}}{2m} \right) + \Delta\overline{\sigma}_\te,\adb  \label{eq:T0}\\
\Delta\overline{\sigma}_\tm &=& \frac{8}{\pi} \Re \int_m^\infty dz \frac{q (q^2 + v_F^2 k^2+ 4m^2) - \lambda r }{r(r + q \lambda)} \left(\frac{1}{e^{\frac{z + \mu}{T}} +1} + \frac{1}{e^{\frac{z - \mu}{T}} +1}\right),\ann
\Delta\overline{\sigma}_\te &=& \frac{8}{\pi \lambda} \Re \int_m^\infty dz \frac{(4m^2+q^2) (\lambda^2 q^2+v_F^2k^2q^2 + 4 m^2 v_F^2 k^2) - \lambda^2 q^2(\lambda^2 +v_F^2 k^2)}{r(\lambda^2q^2 +v_F^2 k^2 q^2+ 4m^2 k^2 v_F^2 + q \lambda r)} \left(\frac{1}{e^{\frac{z + \mu}{T}} +1} + \frac{1}{e^{\frac{z - \mu}{T}} +1}\right), \label{eq:SigmaTot}
\end{eqnarray}
\end{strip}
\noindent where $r = \sqrt{(\lambda^{2}+v_F^2 k^2)^2 (q^2 + k^2 v_F^2) + 4m^2 k^2 v_F^2}$ and $q=\lambda-2i z$. One notes that the first terms in Eqs. \eref{eq:T0} do not contain temperature or chemical potential. These $T$ and $\mu$ dependences are found in the $\Delta\overline{\sigma}_{\te, \tm}$ terms. The \TM\ and \TE\ optical response expressions in Eqs. \eref{eq:T0}, \eref{eq:SigmaTot} take into account the temporal dispersion due to the frequency, spatial dispersion due to the wave vector ${\bf k}$, temperature $T$, chemical potential $\mu$, and finite mass gap $m$. Note that the ${\bf k}$-dependence in the longitudinal (TM) and transverse (TE) conductivity components are different. In the limit of  $k=0$ with $m=\mu=0$ we recover the isotropic and homogeneous case 
\begin{equation}
\overline{\sigma}_\tm = \overline{\sigma}_\te = \frac{8T\ln 2}{\pi \lambda } + \frac{2}{\pi}\int_0^\infty\frac{\tanh(\lambda x/4T)}{x^2+1}dx
\end{equation}
as obtained by others (see Ref. \cite{Gusynin:2007:mocig,Falkovsky:2007:stdogc}). Further taking $T\rightarrow 0$ we also recover the well known universal conductivity value $\overline{\sigma}_\tm = \overline{\sigma}_\te = 1\ (\sigma_\tm = \sigma_\te = \sigma_{gr} = e^2/4)$. Let us also study the 
$T\rightarrow 0$ limit in the case when $m$ and $\mu$ are different from zero with spatial dispersion taken into account. Using Eqs. \eref{eq:T0}, \eref{eq:SigmaTot} we find that when $\mu < m$ the  $\Delta\overline{\sigma}_\tm, \Delta\overline{\sigma}_\te$ terms are zero. When $\mu > m$, however, all contributions to $\overline{\sigma}_\tm, \overline{\sigma}_\te$ in Eq. \eref{eq:SigmaTot} are nonzero.

\begin{figure}[ht]
\centering	\includegraphics[width=9.8truecm]{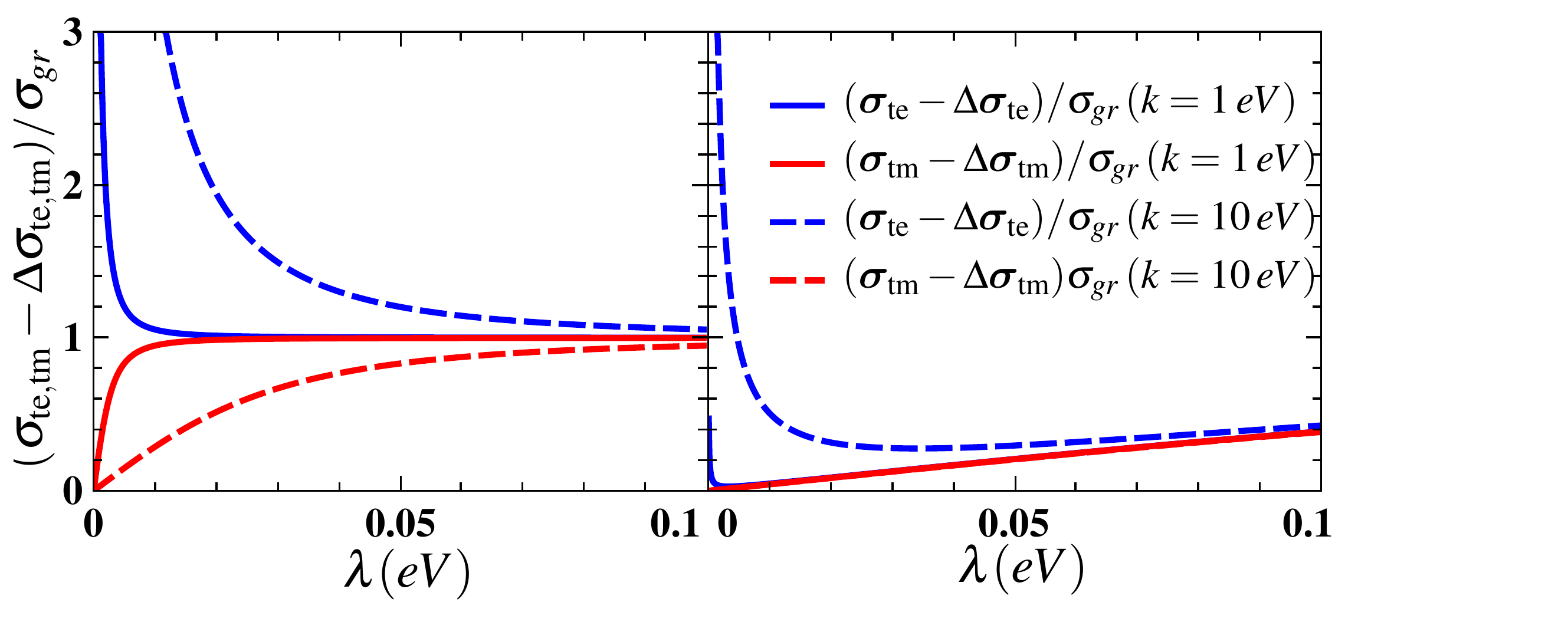}
	\caption{The normalized to $\sigma_{gr} = e^2/4$ graphene conductivity components difference $ \overline{\sigma}_{\tm,\te} - \Delta\overline{\sigma}_{\tm,\te}$ for a mass gap $m=0$ (left panel) and $m=0.1$ eV (right panel) and different values of the wave vector $k$. The line-color legend is the same in both panels. }\label{fig:cnd1}
\end{figure} 

To further analyze the different characteristic dependences in the \TM\ and \TE\ optical response, in Fig.\,\ref{fig:cnd1} we show how the normalized to $\sigma_{gr}$ temperature independent component difference $(\overline{\sigma}_{\tm,\te} - \Delta\overline{\sigma}_{\tm,\te})/\sigma_{gr}$ evolves as a function of the imaginary frequency $\lambda= i \omega$ for different $m$ and $k$ values.  Fig. \ref{fig:cnd1} shows that the low frequency response is dominated by the \TE\ contributions which diverge as $\lambda\rightarrow 0$. The wave vector, however, enhances the role of the \TM\ modes for gapeless graphene as evident from the left panel, where more pronounced region of nonlinear behavior is seen when compared with the $m=0.1$ $eV$ case. As $\lambda$ increases $(\overline{\sigma}_{\tm,\te} - \Delta\overline{\sigma}_{\tm,\te})/\sigma_{gr}$ approaches $1$, which corresponds to the universal graphene conductivity. One notes that this limit is reached much slower for gapped graphene when compared with the $m=0$ case. 

\begin{figure}[ht]
	\centering \includegraphics[width=9.8truecm]{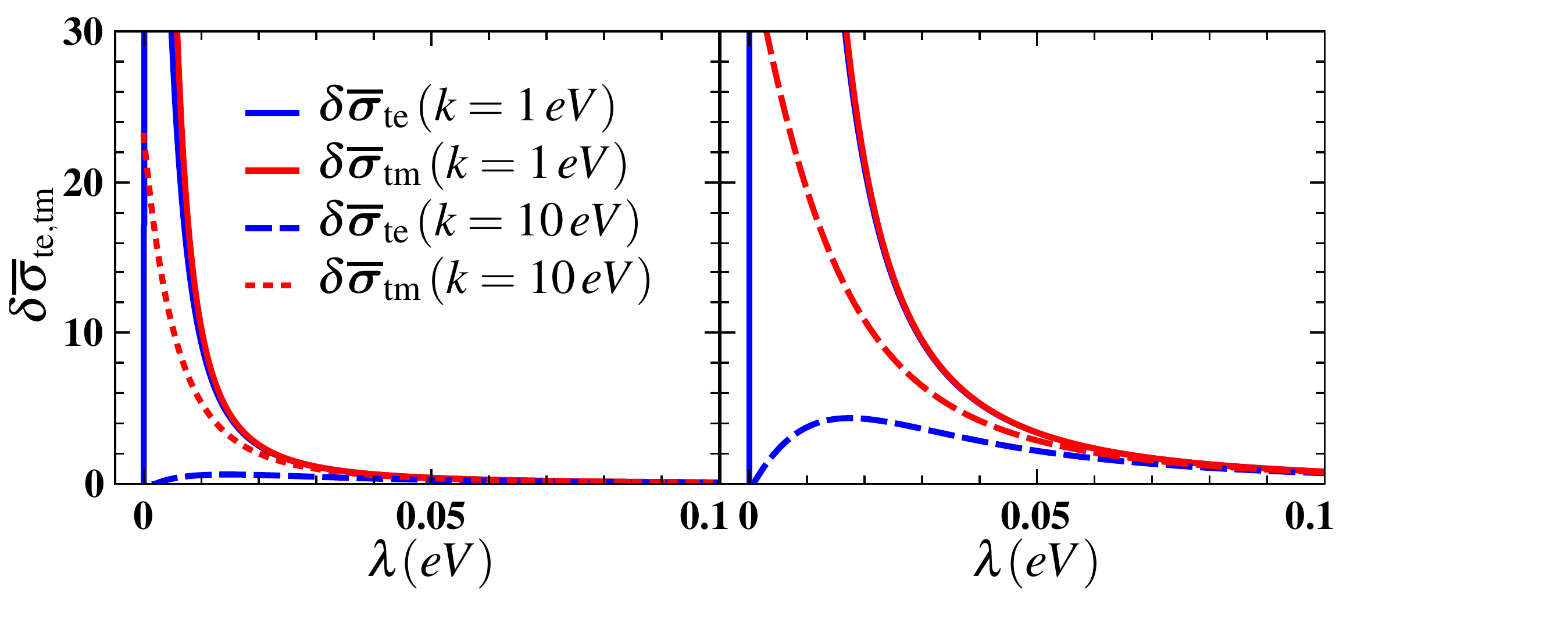}\caption{The ratio $\delta \overline{\sigma}_{\te,\tm} = \Delta \overline{\sigma}_{\te,\tm}/(\overline{\sigma}_{\tm,\te} - \Delta\overline{\sigma}_{\tm,\te})$ at $T = 30$ K (left panel) and $T = 300$ K (right panel) at different values of the wave vector $k$. Here $\mu = 0.1$ eV and $m=0.1$ eV. The line-color legend is the same for both panels. }\label{fig:cnd2}
\end{figure}

The effects of temperature are shown in Fig.\,\ref{fig:cnd2} for a graphene with nonzero chemical potential and mass gaps for different values of the wave vector. One finds that temperature affects primarily the low frequency regime, where the \TE\  contribution is small while the \TM\ part has a large finite value. The effect in \TE\ vs \TM\ disparity is more pronounced for smaller temperature and wave vector values.

\section{Method of Calculations}

\begin{figure}[ht]
\centering \includegraphics[width=4.3truecm]{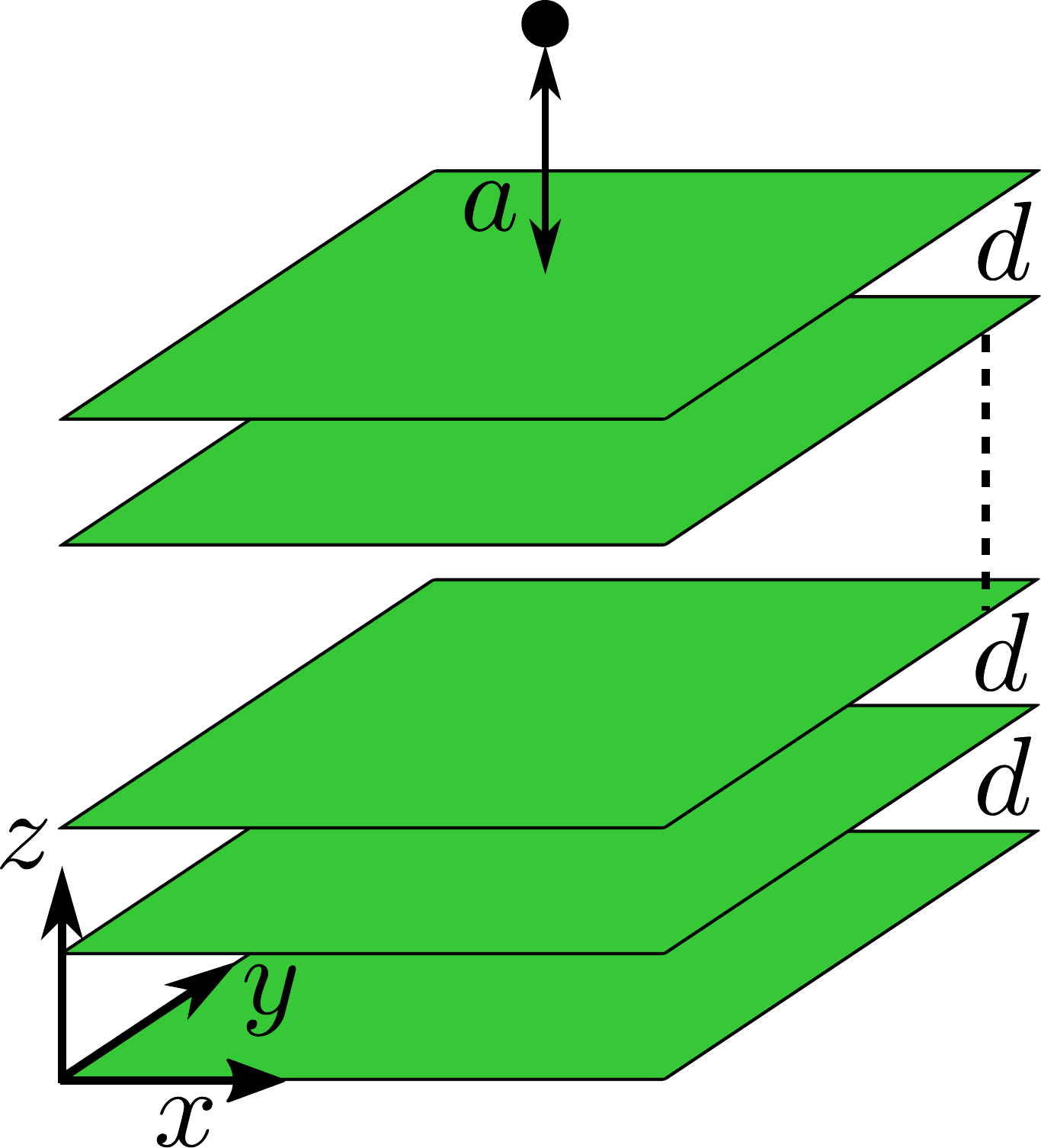}\caption{Schematic representation of infinitely thin layers  equally spaced by a distance $d$ along the $z$-direction. The Casimir-Polder interaction is calculated for an atom placed at a distance $a$ above the top layer, while the Casimir energy for the stack only is calculated using the shown coordinate system. }
\label{fig:stack}
\end{figure}

The system under consideration here consists of $N$ equally spaced infinitely thin layers along the $z$ axis such that each layer extends in the $x-y$ plane as shown in Fig. \ref{fig:stack}. We are interested in the Casimir energy stored in this stack of planes. For the Casimir-Polder interaction initially we take that the half space above the top layer and including the atom is occupied by a dielectric medium specified with a dielectric function $\epsilon(\omega)$, which is then rarefied \cite{Lifshitz:1956:ttomafbs}. As a result, $\epsilon(\omega)=1+4\pi L\alpha(\omega)$, where $L$ is the number of atoms making up the medium and their atomic polarizability is $\alpha(\omega)$. In the limit of $L\rightarrow 0$, one obtains the Casimir-Polder interaction between one atom and the stack of planes, as shown in Fig. \ref{fig:stack}. We find that the Casimir ($\Ca$) free energy per unit area and the Casimir-Polder ($\CP$) free energy can be given as 
\begin{equation}
\cF^{\Ca,(\CP)} = \cF_\tm^{\Ca,(\CP)} + \cF_\te^{\Ca,(\CP)}. 
\end{equation}

The \TE\ and \TM\ contributions are obtained explicitly by using the electromagnetic boundary conditions for the systems in Fig. \ref{fig:stack} and summing the zero-point energy excitations \cite{Khusnutdinov:2016:cpefasocp,Khusnutdinov:2015:cefasocp,Kashapov:2016:tcefpls} 
\begin{eqnarray}
\mathcal{F}_\tm^\Ca &=& \frac{T }{4\pi^2}\sum_{n=0}^\infty{}' \int d^2 k_\perp \ln \Psi_N\left( \frac{\eta_n^\tm  \kappa_n}{\xi_n} \right),\ann
\mathcal{F}_\te^\Ca &=& \frac{T }{4\pi^2}\sum_{n=0}^\infty{}' \int d^2 k_\perp \ln \Psi_N\left( \frac{\eta_n^\te  \xi_n}{\kappa_n}  \right),\label{eq:FCa}\\
\mathcal{F}^{\CP}_\tm &=& \frac{T }{2\pi}\sum_{n=0}^\infty{}'\int d^2 k_\perp  \alpha_n \Phi_N\left( \frac{\eta_n^\tm \kappa_n}{\xi_n} \right)\left(\frac{\xi_n^2}{\kappa_n^2} -2\right),\ann
\mathcal{F}^{\CP}_\te &=& \frac{ T }{2\pi}\sum_{n=0}^\infty{}' \int d^2 k_\perp \alpha_n \Phi_N\left( \frac{\eta_n^\te \xi_n}{\kappa_n}  \right)\left(-\frac{\xi_n^2}{\kappa_n^2}\right),\label{eq:Fa}
\end{eqnarray}
where the following auxiliary functions are defined
\begin{eqnarray}
\Psi_N(t) &=& \frac{e^{-d \kappa_n (N-1)}}{(1+t)^N}\frac{1}{f(t)^{N-1}} \frac{1-f(t)^{2N}}{1-f(t)^2} \ann
&\times& \left(1 + t - e^{-d \kappa_n }f(t) \frac{1-f(t)^{2(N-1)}}{1-f(t)^{2N}}\right),\ann 
\Phi_N(t) &=& \frac{t z e^{-2a \kappa_n }}{1 + t - e^{-d \kappa_n }f(t) \frac{1-f(t)^{2(N-1)}}{1-f(t)^{2N}}}. 
\end{eqnarray}
Here $\kappa_n = \sqrt{k_\perp^2 + \xi_n^2 }$ and $\alpha_n = \alpha(\xi_n), \eta_n^{\te,\tm} = 2\pi \sigma_{\te,\tm} (\xi_n)$ with $\xi_n = 2\pi n  T$ being the Matsubara frequencies. Also, we have defined $f(t) = \sqrt{(\cosh d \kappa_n + t \sinh d\kappa_n)^2-1}+ \cosh d\kappa_n + t \sinh d\kappa_n$. The prime in each summation denotes that the zero term is multiplied by $1/2$.

Eqs. \eref{eq:FCa} and \eref{eq:Fa} constitute the main framework of calculating the Casimir and Casimir-Polder interactions in a multilayered graphene system. The separation into \TE\ and \TM\ contributions achieved in the optical response of an individual graphene is an important factor in the interaction forces being written as a sum of such parts. Let us note that these expressions are quite general as they take into account the distance between the layers, the finite number of layers, and temperature. Additionally, these results also include the spatial dispersion via the 2D wave vector in the graphene conductivity tensor. Potentially, Eqs. \eref{eq:FCa} and \eref{eq:Fa} can be applied to other multilayered materials characterized by different optical conductivity properties. 

\section{Results and Discussion}

Before considering the graphene multilayers, we investigate the limiting case of a stack composed of infinitely conducting planes. We find it convenient to recast Eqs. \eref{eq:FCa} and \eref{eq:Fa} in a form using Poisson's formula 
$\sum_{n=-\infty}^{\infty}\phi(n)=4\pi \sum_{l=0}^{\infty}\int_{0}^{\infty}\phi(s)\cos(2\pi ls)ds$ (see  Ref. \cite{Bordag:2009:ACE} for details). Let us note that the $T=0$ limiting case, found by substituting $T\sum'^\infty_n \to \int_0^{\infty}\frac{d\omega}{2\pi}$ corresponds to the $l=0$ term in the Poisson's expressions for the energies, as reported in Ref. \cite{Khusnutdinov:2016:cpefasocp,Khusnutdinov:2015:cefasocp}. 

Using the Poisson's summation formula, the Casimir interaction in the stack of infinitely conducting layers is obtained essentially as the energy between two  planes with $\sigma\rightarrow \infty$ written as 
\begin{eqnarray}
\mathcal{F}^\Ca &=&\frac{N-1}{\pi^2 d^3}\sum_{l=0}^\infty{}'  \int_{0}^\infty y^2 dy \int_0^1 dx 
\cos \left(\frac{y x l}{dT}\right)   \ann
&\times& \ln \left(1 - e^{-2y} \right).
\end{eqnarray}
The above expression enables us to find the small and large temperature limits,
\begin{eqnarray}
\left. \mathcal{F}^\Ca \right|_{T\to 0} &=& (N-1)\mathcal{E}^\Ca_0  \left\{1 + \frac{45\zeta_R(3)}{\pi^6}\left(2\pi T d\right)^3 \right\},\ann
\left. \mathcal{F}^\Ca \right|_{T\to \infty} &=&  - \frac{\zeta_R(3) T}{8\pi d^2}, \label{eq:Cid}
\end{eqnarray}
where $\zeta_R(3)$ is Riemann zeta function and $\mathcal{E}^\Ca_0 = - \pi^2 /720 d^3$ denotes the quantum mechanical ($T=0$) energy for two infinitely conducting planes multiplied by $(N-1)$. Eqs. \eref{eq:Cid} show that the low $T$ correction to the standard quantum mechanical interaction between the perfectly conducting planes is $\sim T^3$. The high $T$ limit is consistent with the classical thermal fluctuations results for the $n=0$ Matsubara frequency. 

\begin{figure}[ht]
\includegraphics[width=8truecm]{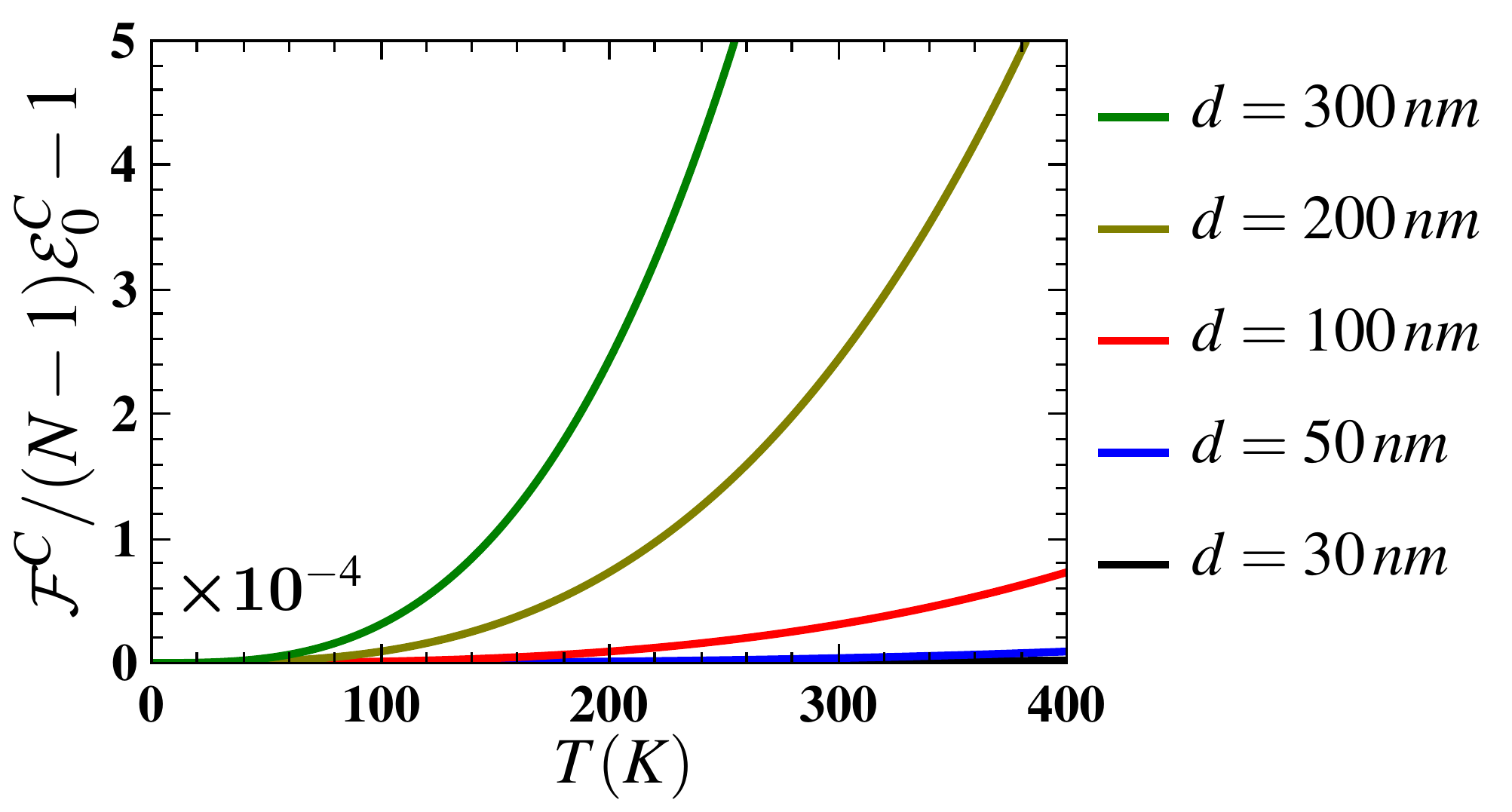}
\includegraphics[width=8truecm]{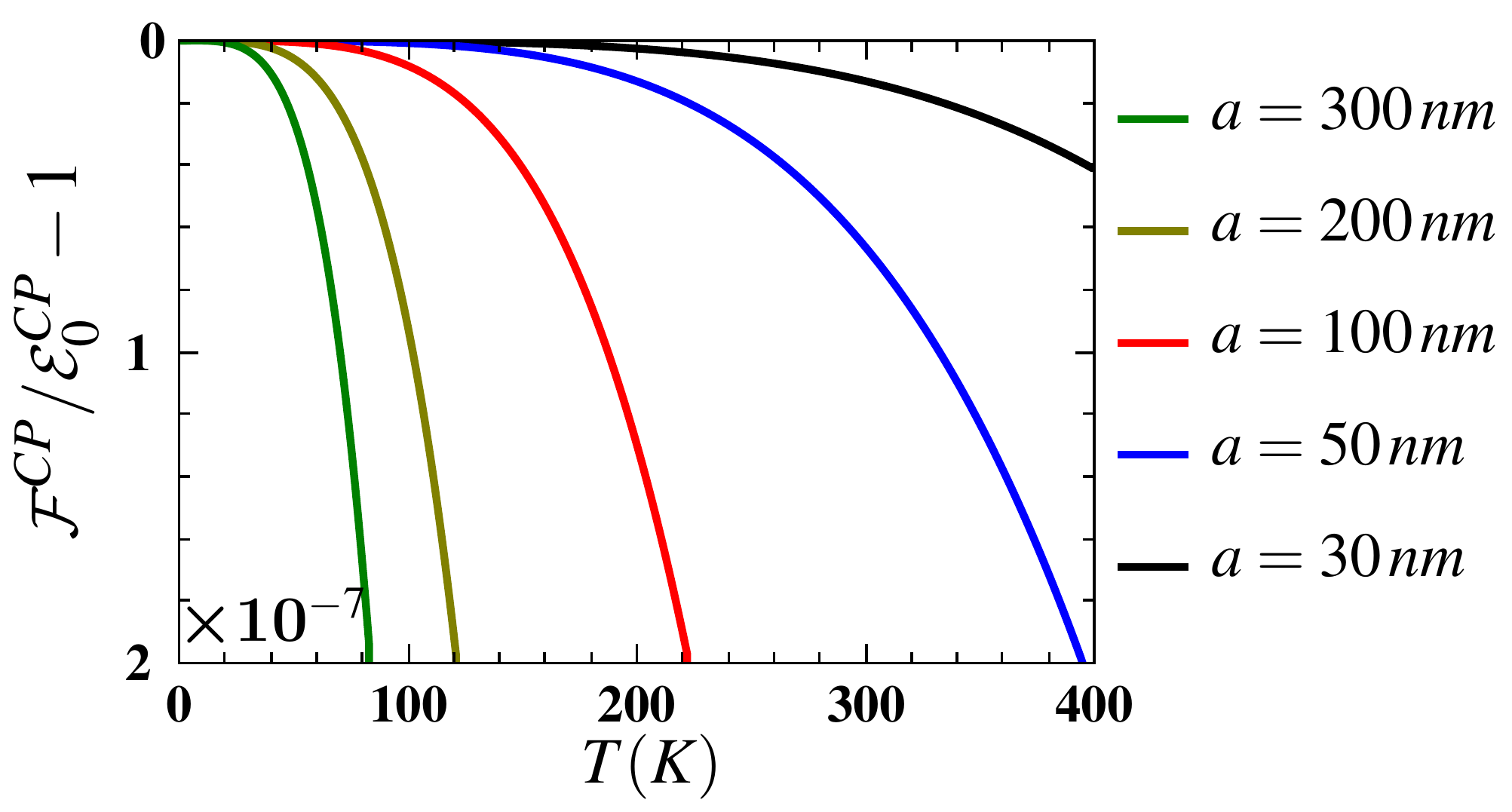}
\caption{(a) The Casimir energy stored in a stack of infinitely conducting planes and normalized to $\mathcal{E}^{\Ca}_0$ for different interplane distances. (b) The Casimir-Polder energy normalized to  $\mathcal{E}_{0}^{\CP}$ between a hydrogen atom and a stack of planes for different atom-stack distances.}\label{fig:CandCPstack}
\end{figure}

\begin{figure*}[htb]
\includegraphics[width=5.8truecm]{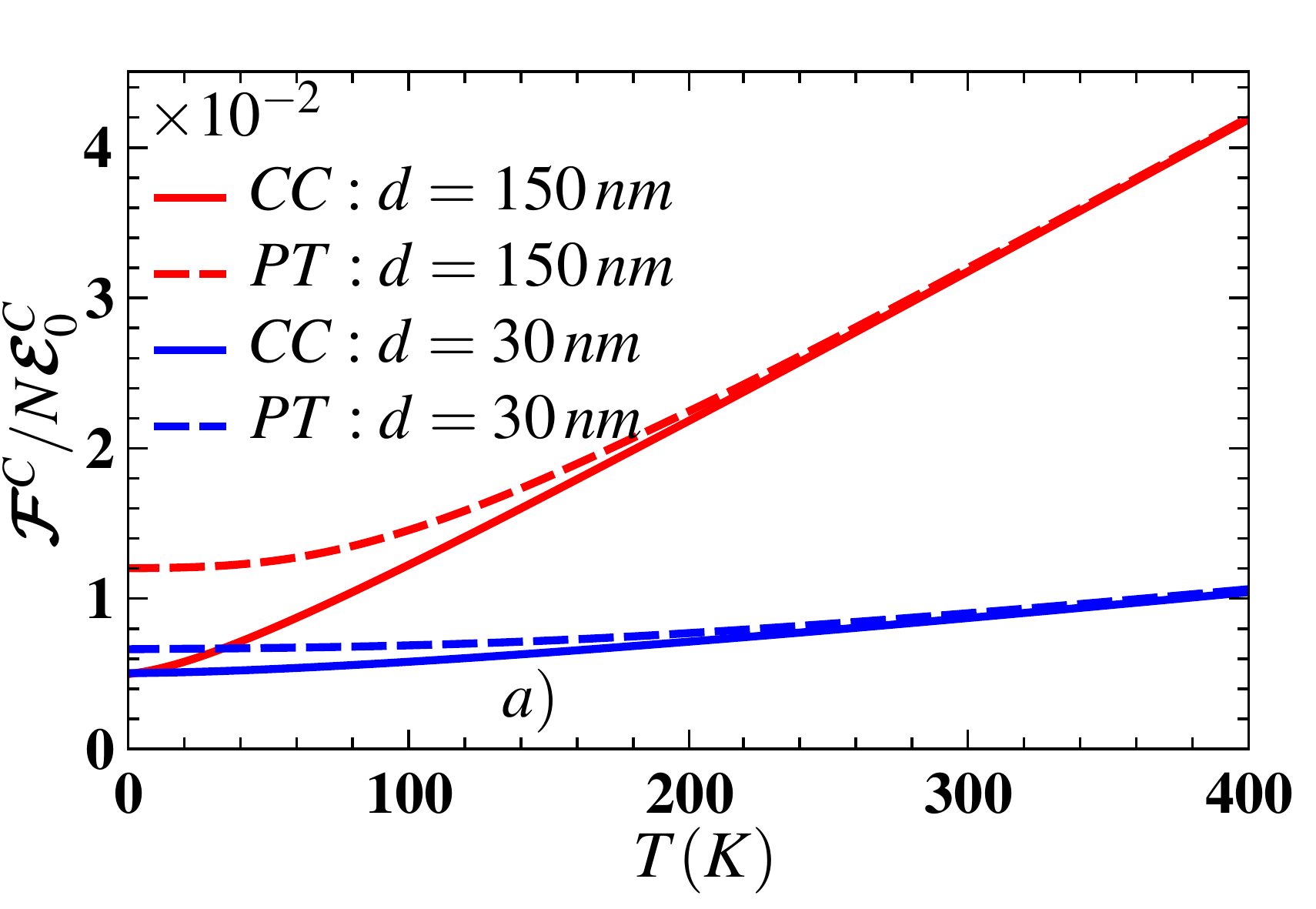}\includegraphics[width=5.8truecm]{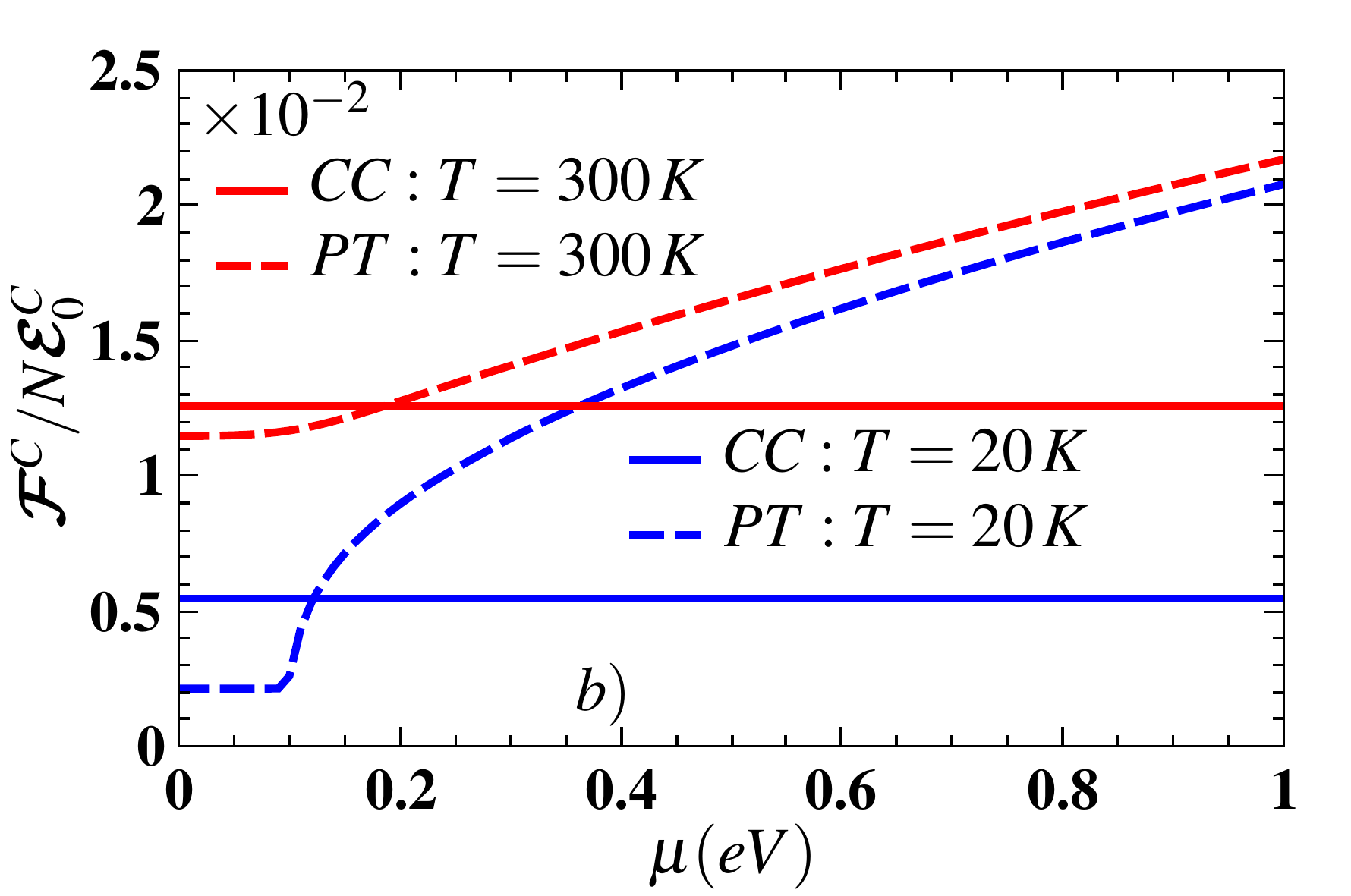}\includegraphics[width=5.8truecm]{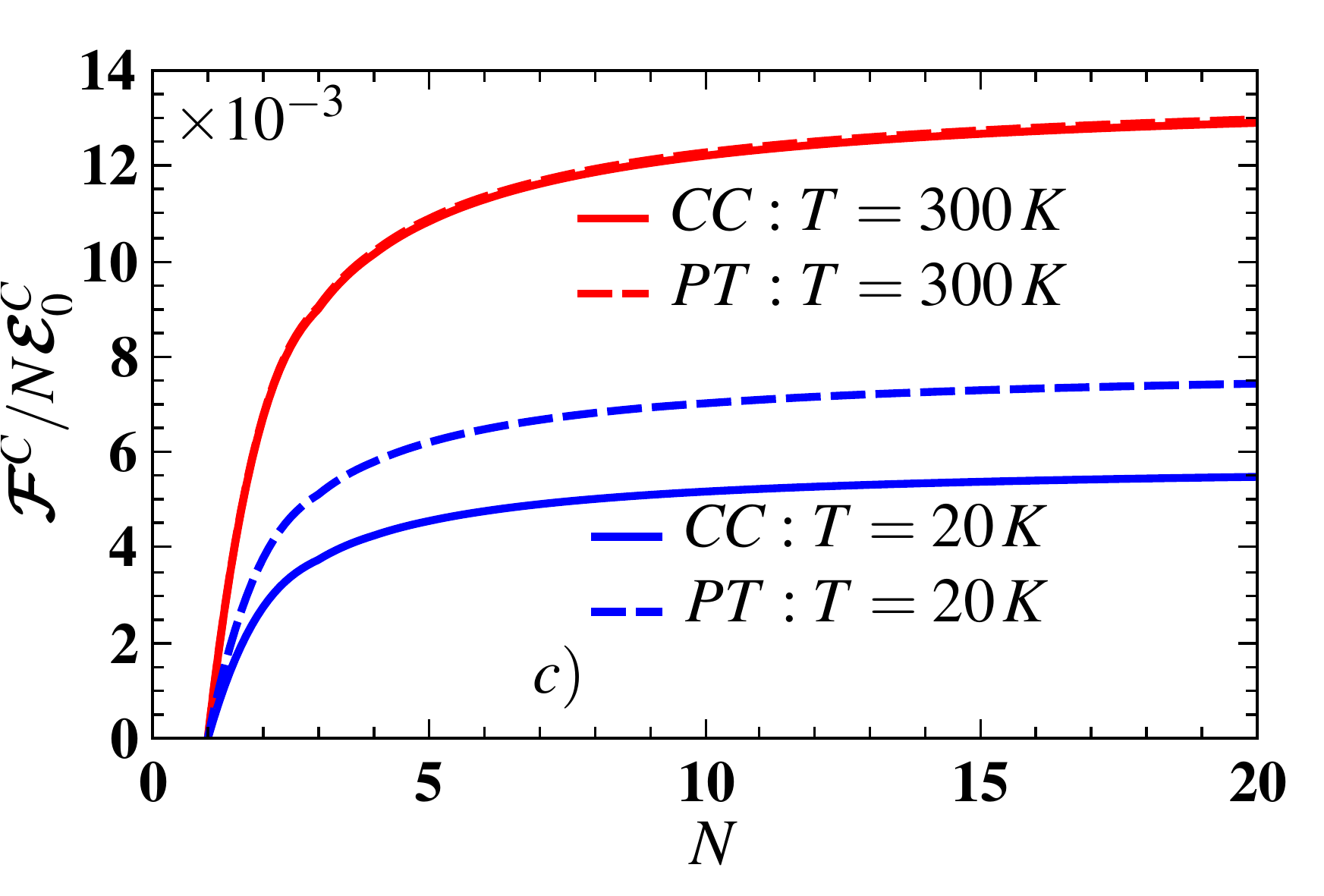}
\caption{The Casimir free energy normalized to the $N\mathcal{E}^\Ca_0$ vs temperature $T$ (panel a), chemical potential $\mu$ (panel b) and number of planes $N$ (panel c) for different models of conductivity (CC -- constant conductivity, solid line; PT -- polarization tensor approach, dashed line). We used $ N=10, m =0, \mu =0.1$ eV (panel a); $N=10, m=0.1$ eV, $d=50$ nm (panel b) and $d=50$ nm,$\mu = 0.1$ eV, $m=0.05$ eV (panel c).} \label{fig:CasfT1}
\end{figure*}

From the Poisson's formula we also find that the Casimir-Polder energy in the $\sigma\rightarrow \infty$ case becomes 
\begin{eqnarray}
\mathcal{F}^\CP &=&-\frac{1}{2\pi a^4}\sum_{l=0}^\infty{}' \int_{0}^\infty dy (2y^2 + 2y + 1) e^{-2y} \ann 
&\times&\cos \left(\frac{yl}{a T}\right)  \alpha\left(\frac{y}{a}\right).
\end{eqnarray}
For large $a$ separations the atomic polarizability becomes $\alpha(y/a)\rightarrow \alpha(0)$, where $\alpha(0)$ is taken at zero frequency. In this case, the integration over $y$ and the summation over $l$ can be performed explicitly,
\begin{equation}
\mathcal{F}^{\CP}\hspace{-1ex} = \frac{\mathcal{E}^{\CP}_0 \chi}{3}  \left( \coth \chi + \chi \csch^2 \chi + \chi^2 \coth \chi \csch^2 \chi \right),\label{eq:CPid}
\end{equation}
where $\chi= 2\pi a T$ and $\mathcal{E}^\CP_0 =-3\alpha(0)/8\pi a^4$ corresponds to the quantum mechanical Casimir-Polder energy between an atom and an infinitely conducting plane. Eq. \eref{eq:CPid} is in agreement with \cite{Bezerra:2008:Ltaiatqr} and it shows that when the atom is far away from the stack, the interaction is determined by the closest to it layer. It is now easy to see that in the limits of small and high temperatures, the Casimir-Polder energy becomes
\begin{eqnarray}
\left.\mathcal{F}^\CP \right|_{T\to 0} &=&  \mathcal{E}^\CP_{0} \left\{1 - \frac{1}{135}\left(2 \pi T a\right)^4 \right\},\ann
\left.\mathcal{F}^\CP \right|_{T\to \infty} &=& \frac{3\alpha (0)}{4a^3} T.\label{eq:idealCP}
\end{eqnarray}
These expressions show that the low $T$ correction to the quantum mechanical result for the energy is $\sim T^4$, which is different that the Casimir-Polder case in Eqs. \eref{eq:Cid}. Similarly to the Casimir interaction, the high temperature limit for the Casimir-Polder energy corresponds to the $n=0$ Matsubara term from Eq. \eref{eq:Fa} describing the thermal fluctuation contribution.

Results for the calculated energies involving infinitely conducting planes are shown in Fig. \ref{fig:CandCPstack}. The left panel indicates that the stored Casimir energy for small separations is not significantly affected by temperature. However, as $d$ is increased, the energy begins to deviate from $\mathcal{E}^\Ca_{0}$, which characterizes the $T=0$ interaction between infinitely conducting objects. The deviations for smaller $T$  appear as $d$ becomes larger as can be seen from the $d=100$ and $300$ nm distances. Similar trends are found for the Casimir-Polder interaction. This behavior is in agreement with the low temperature approximations given by Eq. \eref{eq:CPid} and \eref{eq:Cid}, where correction to the Casimir energy is $\sim T^3$ and Casimir-Polder energy is $\sim (- T^4)$.

Let us now consider the Casimir and Casimir-Polder interactions involving a stack of graphene layers, as specified in Fig. \ref{fig:stack}. Asymptotic low and high temperature expansions are found for the Casimir and Casimir-Polder interactions when the response is taken to be described by the constant graphene universal conductivity, 
\begin{eqnarray}
\left. \mathcal{F}^\Ca \right|_{T\to 0} &=& \mathcal{F}^\Ca_{T=0} \left\{1 + B(\sigma_{gr}, N) \left(2\pi T d\right)^2 \right\},\ann
\left. \mathcal{F}^\Ca \right|_{T\to \infty} &=&  - \frac{\zeta_R(3)}{8\pi d^2} T, \label{eq:Cideal}\adb\\
\left.\mathcal{F}^\CP \right|_{T\to 0} &=& \mathcal{F}^\CP_{T= 0} \left\{1 + A(\sigma_{gr}, N)\left(2\pi T a\right)^2 \right\},\ann
\left.\mathcal{F}^\CP \right|_{T\to \infty} &=& \frac{3\alpha (0)}{4a^3} T.\label{eq:CPn}
\end{eqnarray}
Here $\mathcal{F}^\Ca_{T=0}$ and $\mathcal{F}^\CP_{T= 0}$ are the Casimir and Casimir-Polder energies, respectively,  in the quantum mechanical limit where the Matsubara frequency summation is transformed into an integral ($T\sum_{n=0}^\infty{}' \rightarrow \int\frac{d\omega}{\pi}$) in Eqs. \eref{eq:FCa}, \eref{eq:Fa} and $T=0$ in the optical response in Eq. \eref{eq:T0}, \eref{eq:SigmaTot} \cite{Khusnutdinov:2016:cpefasocp,Khusnutdinov:2015:cefasocp}. Also, $A$ and $B$ are non-trivial functions of $\sigma_{gr}$ and the number of graphene planes in the stack, but they are temperature independent ($A$ and $B$ are not given explicitly here). It appears that the low $T$ behavior is rather different than the low $T$ behavior of a stack of infinitely conducting planes. Specifically, these corrections to the Casimir and Casimir-Polder interactions are $\sim T^2$ and they are positive. For the infinitely conducting planes the low-$T$ Casimir correction is $\sim T^3$ and it is positive, but the low-$T$ Casimir-Polder correction is $\sim T^{4}$ and it is negative.  In general, analytical formulas for the $A$ and $B$ parameters are not possible, however for a single graphene with a constant $\sigma_{gr}$ conductivity, we find that $A = 1/18\eta^\tm$. The high $T$ limit is determined by the $n=0$ Matsubara mode in Eqs. \eref{eq:FCa}, \eref{eq:Fa}. We note that for the constant conductivity response model the arguments of the $\Psi_N$ and $\Phi_N$ functions tend to infinity for the \TM\ mode, while they vanish for the \TE\ mode. As a result the high $T$ interaction is determined by the \TM\ mode giving Eqs. \eref{eq:idealCP} and \eref{eq:Cideal}. When the graphene is described via the polarization tensor, the arguments of the $\Psi_N$ and $\Phi_N$ functions are constant for the zero Matsubara frequency mode. In the $T\rightarrow \infty$ limit, one arrives at the expected classical thermal fluctuations results. 

\begin{figure*}[t]
\includegraphics[width=5.8truecm]{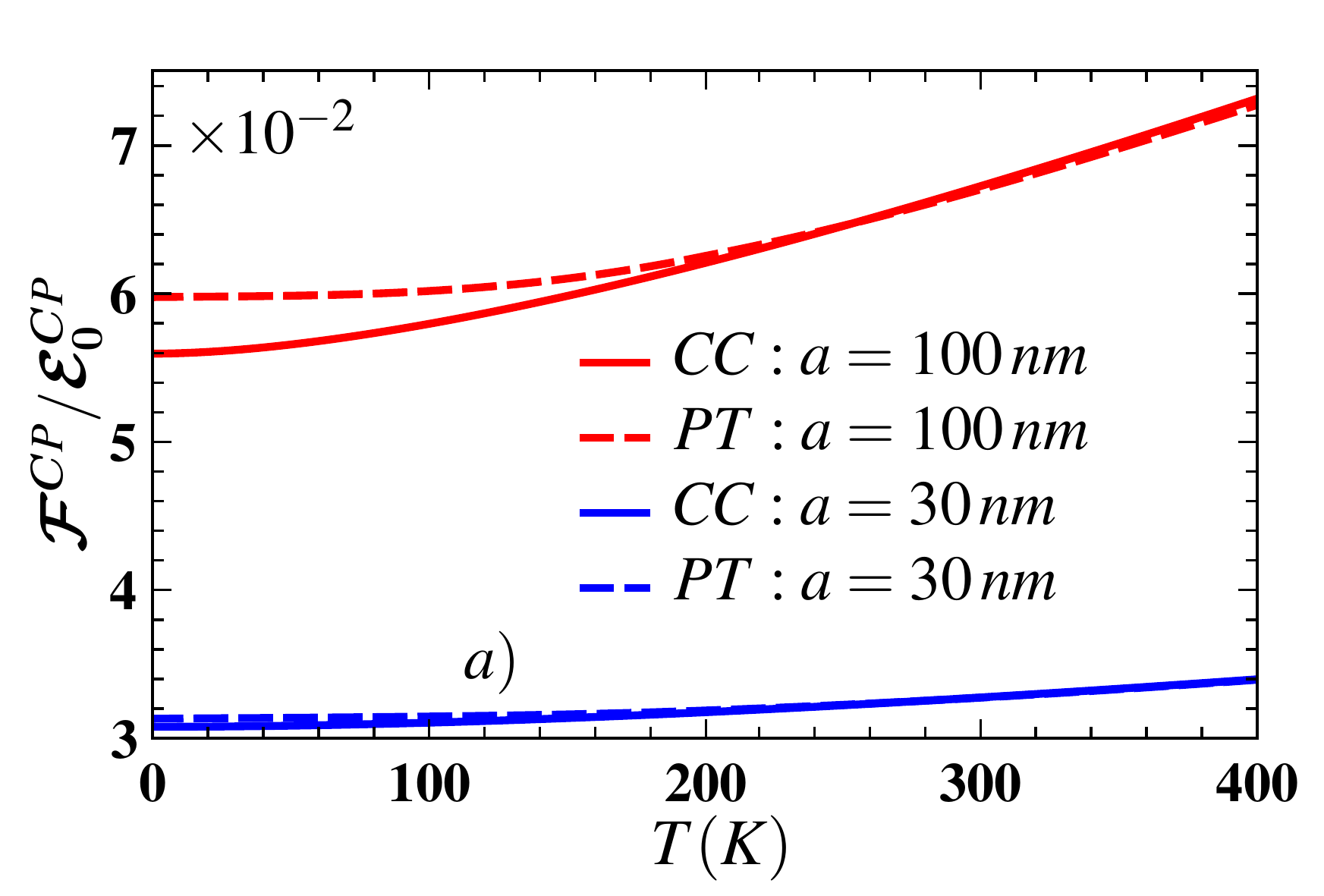}\includegraphics[width=5.8truecm]{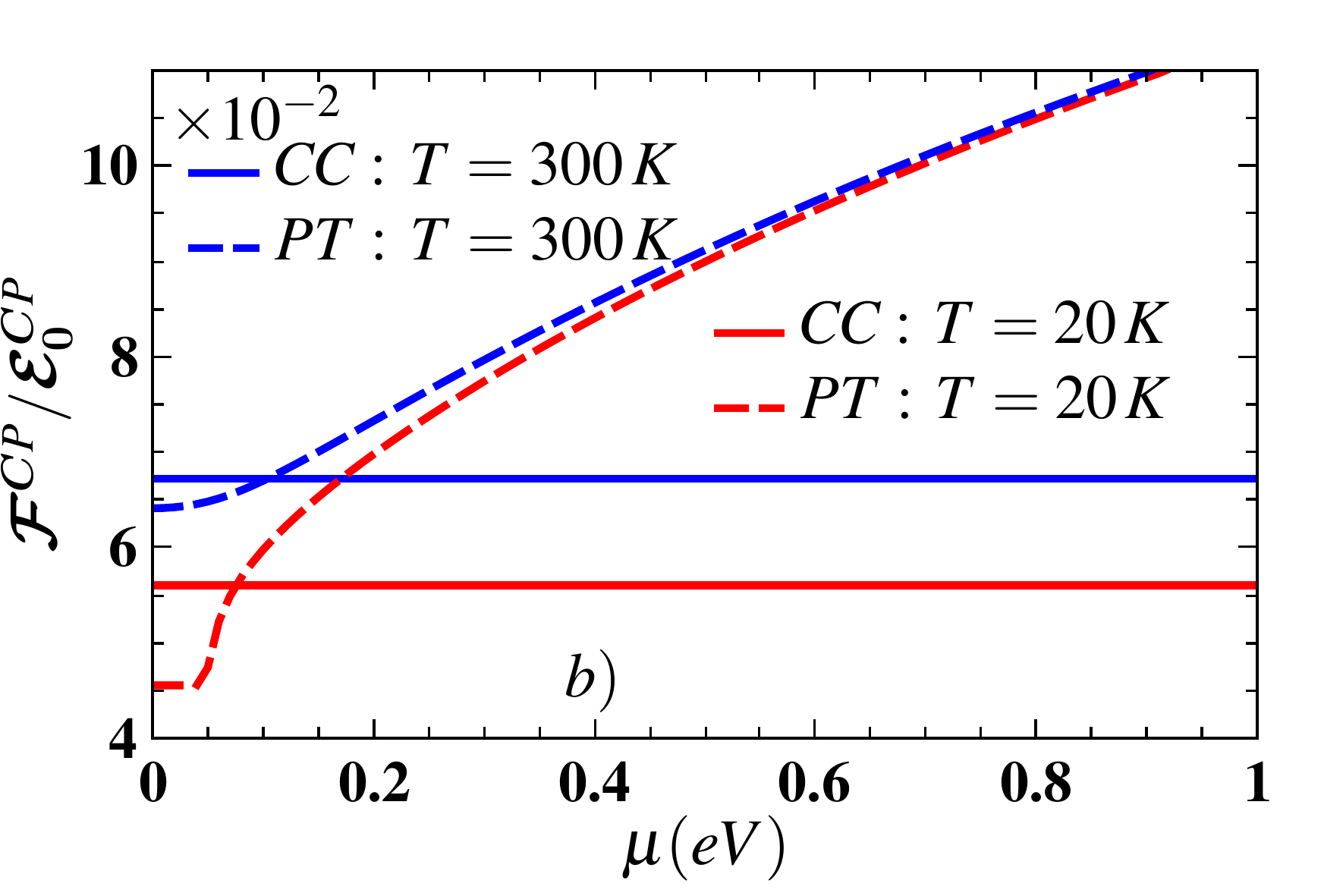}\includegraphics[width=5.8truecm]{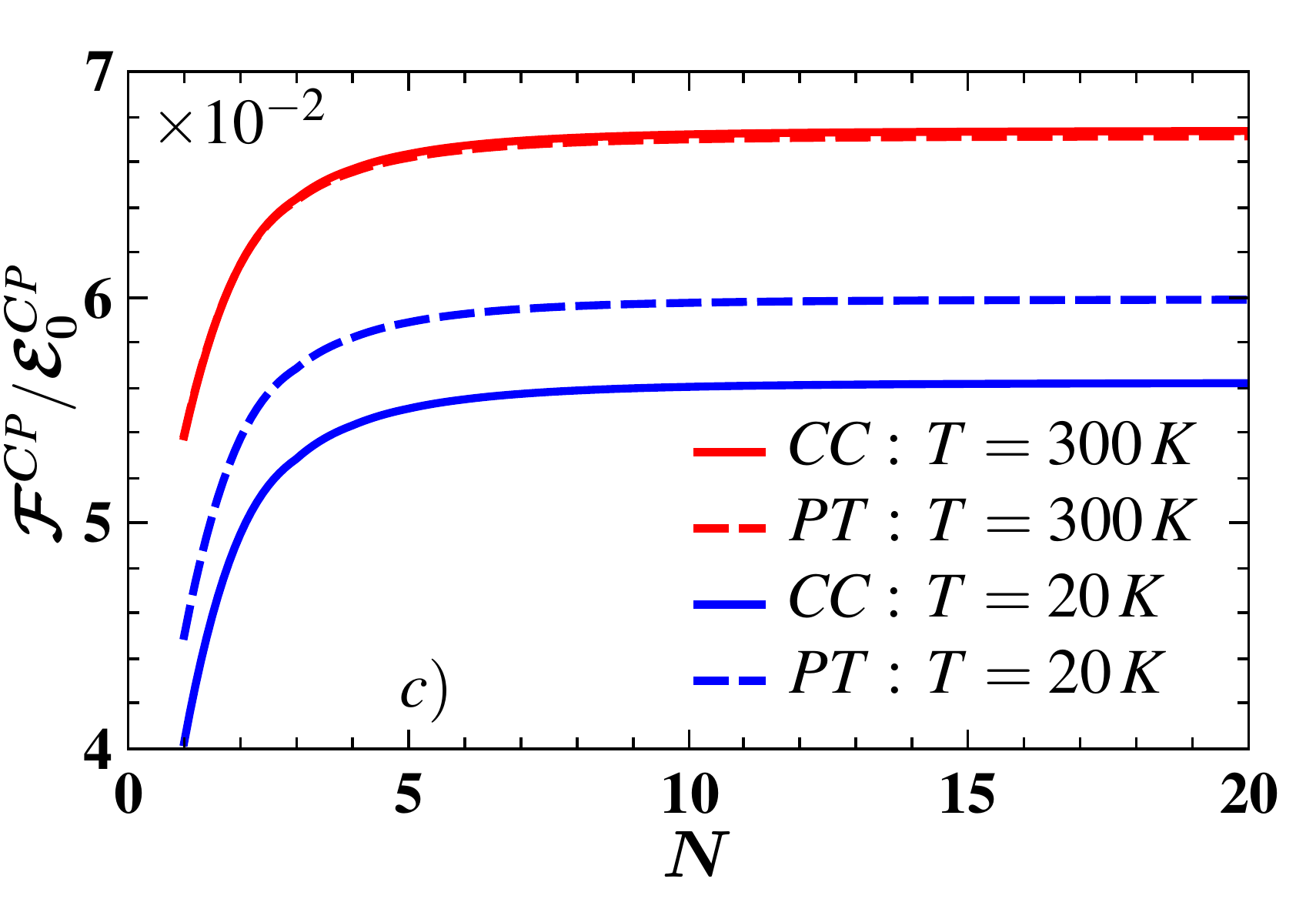}
\caption{The Casimir-Polder free energy normalized to the $\mathcal{E}^\CP_0$ for a Cesium atom above a graphene stack vs temperature $T$ (panel a), chemical potential $\mu$ (panel b) and number of planes $N$ (panel c) for different models of the conductivity (CC -- constant conductivity, solid line; PT -- polarization tensor approach, dashed line). Panel a: chemical potential $\mu =0.1$ eV, mass gap $m=0.05$ eV, interplane distance $d=30$ nm and number of planes $N=10$. Panel b: mass gap $m=0.05$ eV, interplane distance $d=30$ nm, number of planes $N=10$ and distance between atom and stack $a=100$ nm. Panel c: mass gap $m=0.05$ eV, chemical potential $\mu = 0.1$ eV, interplane distance $d=30$ nm and distance between atom and stack $a=100$ nm. Parameters for the Cs atom are taken from \cite{Khusnutdinov:2016:cpefasocp}. } \label{fig:CasPfT1}
\end{figure*}

In Fig.\,\ref{fig:CasfT1} we show numerically calculated results for the Casimir energy stored in a stack of graphene planes by using the constant conductivity and polarization tensor approaches. Panel a) shows that the two models have different results for $T<100$ K for the chosen distances, however, they yield practically the same results for higher temperatures.  This means that the spatial dispersion and frequency dependence taken via the polarization tensor are not important for the Casimir energy in this case. The particular optical response model has a pronounced role when the chemical potential is varied, as depicted in panel b). When $\mu \leq m$ the energy is independent upon the chemical potential, although the polarization tensor model gives smaller values as compared to the ones with $\sigma_{gr}$. As $\mu$ is increased, the intraband transitions (taken into account via the polarization tensor, but not present in $\sigma_{gr}$) begin to dominate and the energy obtained with polarization tensor starts to increase with $\mu$, while the energy obtained with $\sigma_{gr}$ stays constant. Finally, panel c) shows how the interaction depends on the number of graphenes in the stack. For small $N$, the energy is linear with $N$ and it approaches a constant value as the number of planes is increased. The two models for the response essentially give the same result for higher $T$. For smaller temperatures, however, the constant conductivity model underestimates the Casimir energy.

The Casimir-Polder interaction is also investigated numerically by taking a Cs atom above the stack as an example. Fig. \ref{fig:CasPfT1}(a) shows that there are deviations between the constant conductivity and polarization tensor models at smaller temperatures and larger $a$ separations, similar to the situation in Fig. \ref{fig:CasfT1}(a). The behavior of the energy of the Cs atom/graphene stack system as a function of the chemical potential (Fig. \ref{fig:CasPfT1} (b)) and number of planes in the stack ( Fig. \ref{fig:CasPfT1}(c)) is also similar as the one of corresponding Casimir interactions (Fig. \ref{fig:CasfT1}(b), (c)). 

\section{Conclusions}
In this work, a unified description of the Casimir and Casimir-Polder interactions involving a stack of $N$ infinitely thin equally spaced parallel layers is presented. This formalism is applied to graphene and infinitely conducting stacks. Using the planar symmetry separations between TM and TE contributions in the corresponding energies are found. The optical response is key in the interactions and in the graphene case, we consider two optical conductivity models, which involve the constant universal conductivity and the polarization tensor extended over the entire complex frequency plane. Considering various dependences upon separations, chemical potential, and number of planes in the stack and comparing with results for infinitely conducting layers, we show that thermal effects have a unique role in Casimir-like interactions in Dirac-like materials. Thermal fluctuations become strong and even dominant at separations that are much smaller than the typical $\mu m$ scale for standard materials. Our results further indicate that the polarization tensor model for the response is necessary in order to quantify the temperature effects in graphene properly. This study may be useful to experimentalists as new ways to probe classical thermal fluctuations in electromagnetic interactions at the nanoscale.   

The Casimir-Polder energy for two different models of conductivity namely, constant conductivity and conductivity calculated in framework of polarization tensor approach and for atom Cs is shown in Fig. \ref{fig:CasPfT1}.  All parameters for Cs maybe found in Ref. \cite{Khusnutdinov:2016:cpefasocp}.

\ack{
N.K. and R.K. were supported in part by the Russian Foundation for Basic Research Grant No. 16-02-00415-a. N.K. was supported in part by the grants 2016/03319-6 and 2017/50294-1 of S\~ao Paulo Research Foundation (FAPESP). L.M.W. acknowledges financial support from the US Department of Energy under Grant No. DE-FG02-06ER46297. 
}

\section*{References}

\providecommand{\newblock}{}

\end{document}